\documentclass[apjl]{emulateapj}
\usepackage{txfonts}
\usepackage[dvipsnames,usenames]{xcolor}

\newcommand{\code}[1]{\texttt{#1}}

\newcommand*{\maxi}{MAXI~J0556-332}
\newcommand*{\mxb}{MXB~1659-29}

\newcommand*{\xte}{XTE~J1701-462}
                   
\newcommand*{\MeV}{\mathrm{MeV}}

\newcommand*{\gmode}{$\lowercase{g}$-mode}
\newcommand*{\gmodes}{$\lowercase{g}$-modes}

\newcommand*{\mesa}{\code{MESA}}
\newcommand*{\dStar}{\code{dStar}}

\begin{document}

\title{Ocean $\lowercase{g}$-modes on transient neutron stars}
\shorttitle{OCEAN $\lowercase{g}$-MODES ON TRANSIENT NEUTRON STARS}
\author{Alex Deibel\altaffilmark{1,2}}
\affil{
\altaffilmark{1}{Department of Physics and Astronomy, Michigan
State University, East Lansing, MI 48824, USA; deibelal@msu.edu} \\
\altaffilmark{2}{The Joint Institute for Nuclear Astrophysics - Center for the Evolution of the Elements, Michigan State University, East Lansing, MI 48824, USA}  }

\shortauthors{ALEX DEIBEL}

%\email{$^{\star}$deibelal@msu.edu}

\begin{abstract}
The neutron star ocean is a plasma of ions and electrons that extends from the base of the neutron star's envelope to a depth where the plasma crystallizes into a solid crust. During an accretion outburst in an X-ray transient, material accumulates in the envelope of the neutron star primary. This accumulation compresses the neutron star's outer layers and induces nuclear reactions in the ocean and crust. Accretion-driven heating raises the ocean's temperature and increases the frequencies of \gmodes \ in the ocean; when accretion halts, the ocean cools and ocean \gmode \ frequencies decrease. If the observed low-frequency quasi-periodic oscillations on accreting neutron stars are \gmodes \ in the ocean, the observed quasi-periodic oscillation frequencies will increase during outburst\textrm{---}reaching a maximum when the ocean temperature reaches steady state \textrm{---} and subsequently decrease during quiescence. For time-averaged accretion rates during outbursts between $\langle \dot{M} \rangle = 0.1 \textrm{--} 1.0\, \dot{M}_{\rm Edd}$ the predicted \gmode \ fundamental $n=1$ $l=2$ frequency is between $\approx 3 \textrm{--} 7 \, \mathrm{Hz}$ for slowly rotating neutron stars. Accreting neutron stars that require extra shallow heating, such as the Z-sources \maxi, \mxb, and \xte, have predicted \gmode \ fundamental frequencies between $\approx 3 \textrm{--} 16 \, \mathrm{Hz}$. Therefore, observations of low-frequency quasi-periodic oscillations between $\approx 8 \textrm{--} 16\, \mathrm{Hz}$ in these sources, or in other transients that require shallow heating, will support a \gmode \ origin for the observed quasi-periodic oscillations. 
\end{abstract}


\section{Introduction}

The Z-sources are the most luminous neutron star X-ray transients ($L_{\mathrm{X}} \sim 0.5 \textrm{--} 1.0 \, {L_{\rm X, Edd}}$), with inferred mass accretion rates on the order of the Eddington accretion rate, as determined from the X-ray flux. These sources trace out a Z-shaped track in the X-ray color-color diagram \citep{hasinger1989}; the track is composed of the horizontal (top), normal (diagonal), and flaring (bottom) branches. During an accretion outburst, a source may vary its spectral state (its location on the ``Z") over the course of hours or days due to accretion instabilities at a nearly constant accretion rate \citep{lin2009,homan2010,fridriksson2015}. Although changes in spectral state were thought to result from variations in the mass accretion rate, it is likely that accretion rate variations cause secular evolution of the source; for example, \xte \ transitioned into a lower luminosity atoll phase ($L_{\mathrm{X}} \sim 0.1 \textrm{--} 0.5 \, \mathrm{L_{\rm X, Edd}}$) when its accretion rate decreased \citep{lin2009,homan2010}.

On the normal branch, low-frequency quasi-periodic oscillations between $\approx 5\textrm{--}7\, \mathrm{Hz}$ are observed in several Z-sources, for example, Cyg~X-2 \citep{hasinger1989,wijnands1997,dubus2004}, SCO~X-1 \citep{hertz1992, vanderklis1996, titarchuk2014}, and GX~5-1 \citep{kuulkers1994, jonker2002}. The normal branch oscillation frequencies (hereafter NBOs) are consistent with \gmodes \ \citep{bildsten1995,bildsten1996} in the neutron star's ocean \textrm{---} a plasma of ions and electrons that extends from the overlying envelope to the deeper crust \textrm{---} provided that the neutron star is slowly rotating ($\nu_{\rm spin} \lesssim 10 \, \mathrm{Hz}$). Low-frequency quasi-periodic oscillations are also consistent with being of a geometric origin \citep{stella1998,homan2012,homan2015} from Lense--Thirring precession of an inner accretion disk \citep{lense1918,bardeen1975}; for example, horizontal branch oscillation frequencies are consistent with a warped accretion disk geometry \citep{jonker2002}. It has been difficult, however, to delineate the origin of the low-frequency NBOs given the uncertainties that remain in the nature of the accretion flow near the neutron star surface. 

The thermal evolution of the neutron star ocean offers an opportunity to test for the presence of ocean \gmodes \ by comparing the temporal evolution of predicted \gmodes \ and observed NBOs. During an accretion outburst, the accumulation and compression of matter in the neutron star's envelope induces nuclear reactions in the ocean and crust \citep{bisnovatyi1979,sato1979} that deposit $\approx 1\textrm{--}2 \, \mathrm{MeV}$ per accreted nucleon \citep{haensel1990}. As the ocean's temperature increases away from thermal equilibrium with the core, the predicted ocean \gmode \ frequencies also increase. When accretion ceases, the ocean cools back toward thermal equilibrium with the core and predicted ocean \gmode \ frequencies decrease. For example, in sources with time-averaged accretion rates between $\langle \dot M \rangle= 0.5 \textrm{--}1.0 \, \dot{M}_{\rm Edd}$, the predicted fundamental \gmode \ frequency shifts from an initial value of $\approx 3 \, \mathrm{Hz}$ in a cold ocean to $\approx 5\textrm{--}7 \, \mathrm{Hz}$ when the ocean reaches steady state near $\approx 160 \, \mathrm{days}$ after accretion begins \textrm{---} NBO frequencies in the $\approx 5\textrm{--}7 \, \mathrm{Hz}$ range are observed within the same time frame, for example, in \xte \ \citep{homan2007,homan2010}.

Models of neutron star thermal relaxation naturally reproduce quiescent light curves of neutron star transients (e.g., Brown \& Cumming 2009). As determined from their quiescent light curves, some neutron star transients require extra ``shallow" heating during outburst from an unknown source to explain their high temperatures at the outset of quiescence \citep{brown2009, deibel2015, turlione2015}. A finite amount of shallow heating, for instance, the $\approx 1 \, \mathrm{MeV}$ per accreted nucleon needed in \mxb, raises the ocean's steady-state temperature and predicted \gmode \ frequencies may be $\gtrsim 7 \, \mathrm{Hz}$ in this source.  For this reason, any observed NBO frequencies $\gtrsim 7 \, \mathrm{Hz}$ in Z-sources that require extra shallow heating, such as \maxi \ \citep{deibel2015}, \mxb \ \citep{brown2009}, and \xte \ \citep{turlione2015}, would provide an observational test for the presence of ocean \gmodes. These sources have the largest predicted \gmode\ frequencies during outburst, reaching frequencies between $\approx 8\textrm{--} 16 \, \mathrm{Hz}$. Furthermore, because the neutron star's quiescent luminosity is much smaller than the heating rate during outburst in these sources, the large ocean temperatures and \gmode \ frequencies remain for thousands of days into quiescence.

In this study, we examine the thermal evolution of Z-sources and the temporal evolution of the ocean's predicted fundamental \gmode \ in \maxi, \mxb, \xte, and accreting neutron stars with accretion rates similar to the transient Z-sources. In Section~\ref{section.model}, we outline our neutron star thermal evolution model and demonstrate the evolution of the ocean during an accretion outburst. In Section~\ref{section.gmode}, we explore changes in the \gmode \ spectrum over the course of the modeled outbursts. We discuss our results in Section~\ref{section.discussion}. 

\section{The thermal evolution of the ocean} \label{section.model}

We follow the thermal evolution of the neutron star's outer layers by solving the fully general relativistic heat diffusion equation using the open source code \dStar\ \citep{dstar}. The microphysics follow \citet{brown2009} with changes outlined in \citet{deibel2015}. \dStar \ uses the numerics package available as part of the open source stellar evolution code \mesa \ \citep{paxton2011, paxton2013, paxton2015}. Thermal evolution models have successfully modeled quiescent light curves for neutron star transients; for instance, EXO~0748-676 \citep{degenaar09, degenaar2014} and Swift~J174805.3-244637 \citep{degenaar2015}. 

Some sources require additional heating during outburst to reconcile models with quiescent observations, for example, KS 1731-260 \citep{wijnands2001, wijnands2002, cackett2010}  and MXB 1659-29 \citep{wijnands2003, wijnands2004, cackett2008} require $Q_{\rm shallow} \approx 1 \, \mathrm{MeV}$ per accreted nucleon of extra heating in the shallow crust \citep{brown2009}. A recent fit to the quiescent light curve of \xte \ \citep{fridriksson2010, fridriksson2011} requires a $Q_{\rm shallow} \approx 0.2 \, \mathrm{MeV}$ per accreted nucleon heat source in its shallow crust \citep{turlione2015}. The hottest neutron star transient, \maxi \ \citep{homan2011,homan2014, matsumura2011, sugizaki2013}, requires a $Q_{\rm shallow} \approx 4 \textrm{--} 16 \, \mathrm{MeV}$ per accreted nucleon heat source in its shallow crust to fit quiescent observations \citep{deibel2015}. 

\begin{figure}
\centering
\includegraphics[width=1.0\columnwidth]{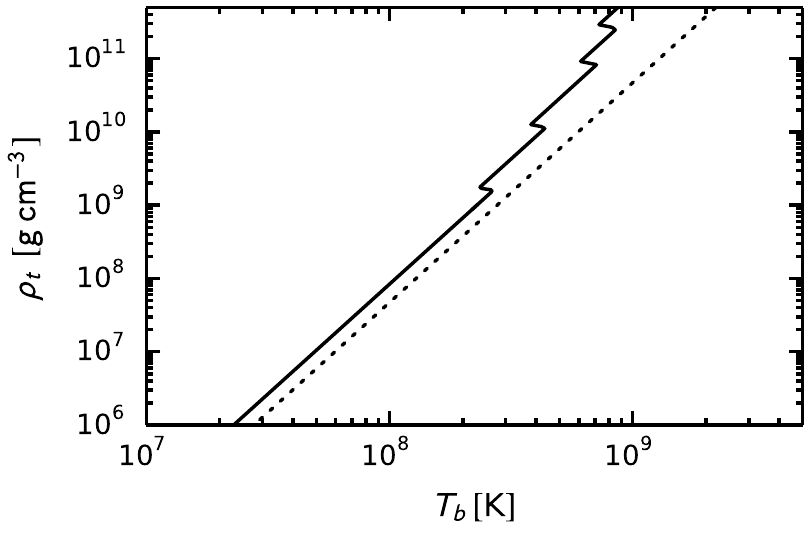}
\vspace{-0.3cm}
\caption{Location of the ocean-crust transition as a function of temperature at the base of the ocean. The black solid curve is for the one-component accreted composition \citep{haensel1990} and the black dotted curve is for a multi-component accreted composition \citep{steiner2012}. The crystallization density is given by Equation~(2). \label{fig:rho_t}
}
\end{figure}

Before the onset of accretion, the neutron star ocean is in thermal equilibrium with the core near $T_{\rm core} \sim 10^{7} \, \mathrm{K}$, as inferred from cooling neutron star transients  \citep{brown1998b,brown2009,page2013}. During active accretion, compression-induced nuclear reactions deposit $\approx 1 \textrm{--} 2 \, \mathrm{MeV}$ per accreted nucleon in the ocean and crust \citep{haensel1990, haensel2003, gupta2007, haensel2008} and the initially cold ocean can reach temperatures of $T_b \gtrsim 10^{8} \, \mathrm{K}$. Moreover, in sources that require shallow heating, the ocean temperature can reach $T_b \gtrsim 10^{9} \, \mathrm{K}$ in steady state \citep{deibel2015}. An increase in the temperature at the base of the ocean increases the crystallization depth of the ocean. The crystallization of the one-component ocean, with proton number $Z$, baryon number $A$, and ion number density $n_i$, depends on the plasma coupling parameter  
\begin{equation}
\Gamma = \frac{(Ze)^2}{ak_{\rm B} T_b} \ ,
\end{equation}

\noindent where $T_b$ is the temperature at the base of the ocean, $a=(3/4 \pi n_i)^{1/3}$ is the inter-ionic spacing, and the crystallization point occurs where $\Gamma = 175$ \citep{farouki93, potekhin2000}. Therefore, the ocean-crust transition occurs at a mass density of
\begin{equation}
\rho_t \approx 2.2 \times 10^{6} \, \mathrm{g \ cm^{-3}} \left(\frac{T_b}{3 \times 10^{7} \, \mathrm{K}} \right)^3 \left(\frac{26}{Z} \right)^6 \left(\frac{A}{56} \right) \ .
\end{equation}
The ocean-crust transition density as a function of temperature can be seen in Figure~\ref{fig:rho_t}. For simplicity, we use a one-component accreted composition \citep{haensel1990} that roughly traces models of the multi-component accreted composition \citep{steiner2012}, as can be seen in Figure~\ref{fig:rho_t}. Note that as the crust liquifies, composition changes in the solid crust become part of the liquid ocean. Sharp composition changes in the ocean, such as those arising from $e^{-}$-capture layers, can trap modes in the ocean \citep{bildsten1998} and decrease mode frequencies \citep{strohmayer1993}. We assume, however, that composition changes do not affect ocean oscillations because a realistic accreted composition contains smooth gradients in $\langle Z \rangle$ and $\langle A \rangle$  \citep{steiner2012}.

\begin{figure}
\centering
\includegraphics[width=1.0\columnwidth]{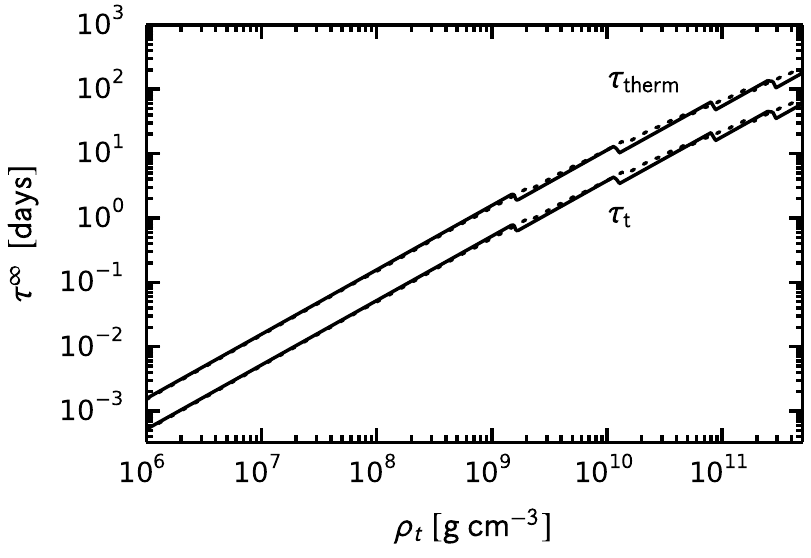}
\vspace{-0.3cm}
\caption{Characteristic timescales in the neutron star ocean. The upper curves are the thermal time for an equilibrium composition (solid curve) and accreted composition (dotted curve). The lower curves are for the timescale for changes in $\rho_t$ for an equilibrium composition (solid curve) and an accreted composition (dotted curve). \label{fig:thermal_time}
}
\end{figure}

The timescale for changes to the ocean-crust transition density is related to the thermal time at the base of the ocean. We use the thermal time from Deibel et al. (2015; their Equation~(1)), 
 \begin{equation} \label{equation.ocean_time}
\tau_{\rm therm, ocean}^{\infty} \approx 1.0\, {\rm days} \ \rho_{9} \ \left(\frac{g_{14}}{2}\right)^{-2} \left({Y_e\over 0.4}\right)^3\left({Z\over 26}\right) \left({1+z\over 1.31}\right) \ ,
 \end{equation}
 
\noindent where $Y_e$ is the electron fraction, $g_{14} \equiv g/10^{14} \, {\rm cm \ s^{-2}}$ is the gravitational acceleration with $g=(1+z)GM/R^2$, $1+z = (1-2GM/(Rc^2))^{-1/2}$ redshifts to an observer frame at infinity, and $\rho_9 \equiv \rho/(10^9 \, \mathrm{g \ cm^{-3}}$). The thermal time in the ocean as a function of mass density can be seen in Figure~\ref{fig:thermal_time}. The timescale for changes to the ocean-crust transition density is $\tau^{\infty}_t = (d\mathrm{ln}T/d\mathrm{ln}\rho_t)(dt/d\mathrm{ln}T) = \tau^{\infty}_{\rm therm,ocean}/3$. 

\maxi \ is the hottest quiescent Z-source to date, and to reproduce the observed MAXI~J0556-332 outburst beginning in 2011 \citep{matsumura2011} we run a model that accretes at $\langle \dot{M} \rangle = 1.0\, \dot{M}_{\rm Edd}$ for 16 months. We use the model parameters from the best fit to the quiescent light curve from \citet{deibel2015}: a neutron star mass $M= 1.5\, \mathrm{M_{\odot}}$, a neutron star radius $R= 11\,{\rm km}$, a core temperature $T_{\rm core}= 3 \times 10^7 \, \mathrm{K}$, an impurity parameter $Q_{\rm imp} = 1$, and a shallow heat source of $Q_{\rm shallow} = 6 \, {\rm MeV}$ per accreted nucleon. We assume the shallow heating varies proportionally with the accretion rate and is deposited over the column depth range $y = 2 \times 10^{13} \textrm{--} 2 \times 10^{14} \, \mathrm{g \ cm^{-2}}$ as inferred by the quiescent light curve \citep{deibel2015}, where $y = P/g$ is the column depth. 

\begin{figure}
\centering
\includegraphics[width=1.0\columnwidth]{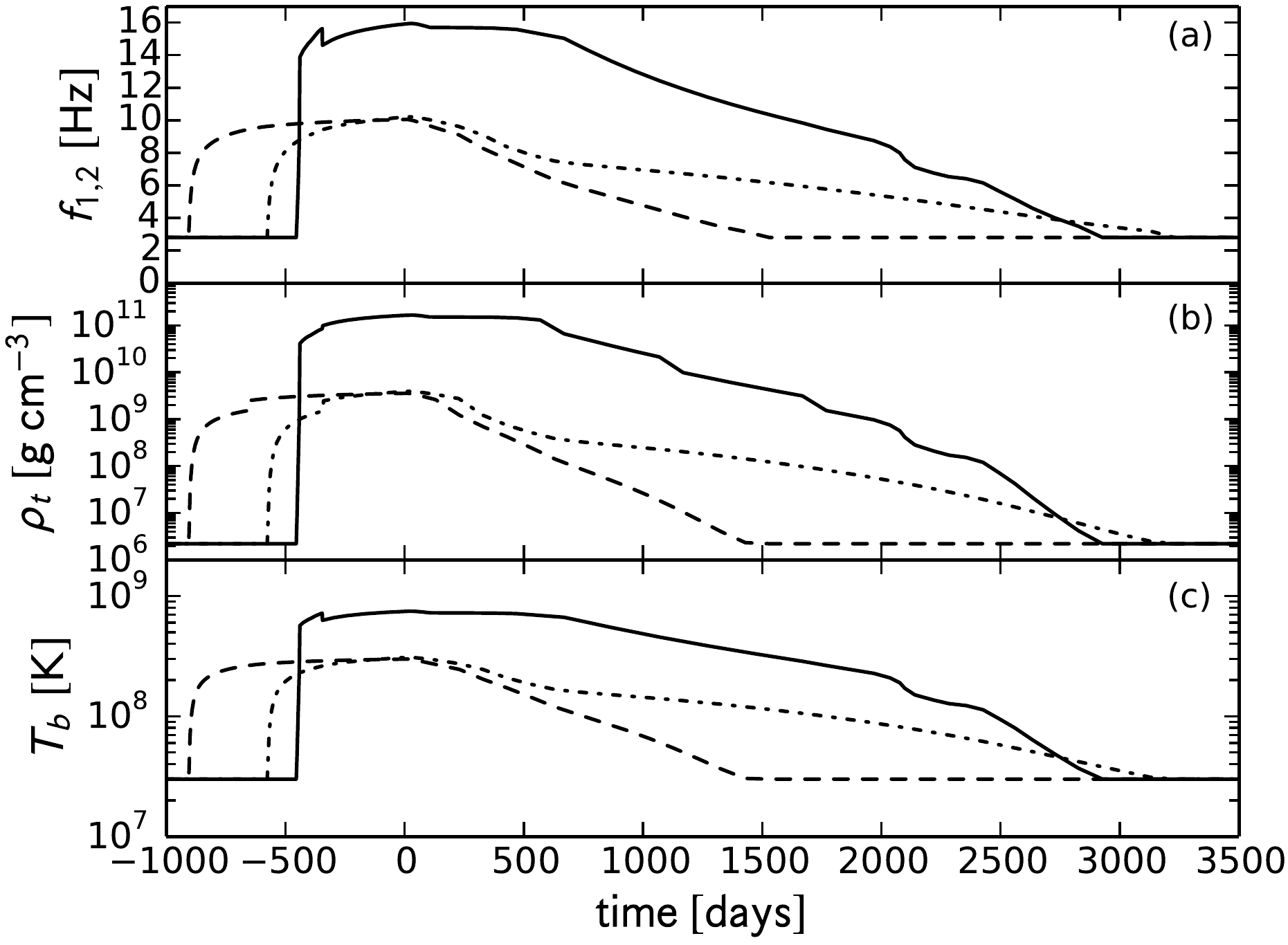}
\vspace{-0.3cm}
\caption{Thermal evolution as a function of time during outburst/quiescence for \maxi \ (solid curves), \mxb\ (dashed curves), and \xte \ (dotted-dashed curves). {Panel (a)}: fundamental $n=1$, $l=2$ \gmode \ frequencies in the ocean (Equation~4). {Panel (b)}: the ocean-crust transition density (Equation~2). {Panel (c)}: temperature at the ocean-crust transition. \label{figure.maxi}}
\end{figure}

In the model of the \maxi \ accretion outburst, the ocean temperature reaches $T_b \approx 7.5 \times 10^{8} \, \mathrm{K}$ in steady state and the ocean-crust boundary moves from $\rho_t \approx 2.2 \times 10^{6}\, \mathrm{g \ cm^{-3}}$ to $\rho_t \approx 1.7 \times 10^{11}\, \mathrm{g \ cm^{-3}}$, as can be seen in Figure~\ref{figure.maxi}. Note that the ocean temperature increases relatively quickly during the outburst due to the combined heating from accretion-driven nuclear reactions and shallow heating, which deposit $L_{\rm nuc} \approx 1.5 \times 10^{37} \, \mathrm{ergs \ s^{-1}}$. The quiescent luminosity of the neutron star, however, is only $L_{\rm q} \approx 2.3 \times 10^{35} \, \mathrm{ergs \ s^{-1}}$, and the neutron star therefore cools over a period much longer than the outburst duration. As a consequence, the ocean remains $\gtrsim 7 \times 10^{8} \, \mathrm{K}$ for $\approx 500 \, \mathrm{days}$ into quiescence and takes $\approx 2900 \, \mathrm{days}$ to cool completely. 

Motivated by renewed activity in \mxb \ \citep{negoro2015}, we explore its thermal evolution in a model representative of its previous $2.5 \, \mathrm{year}$ outburst \citep{wijnands2003,wijnands2004} near $\langle \dot{M} \rangle = 0.1\,\dot{M}_{\rm Edd}$. The model uses a neutron star mass of $M=1.6\, \mathrm{M_{\odot}}$, a neutron star radius of $R=11.2\,{\rm km}$, a core temperature of $T_{\rm core}=3 \times 10^7 \, \mathrm{K}$, an impurity parameter of $Q_{\rm imp} = 4$, and a $Q_{\rm shallow} = 1 \, \MeV$ per accreted nucleon shallow heat source, which agrees with the quiescent light curve fit from \citet{brown2009}. During the accretion outburst, the ocean temperature reaches $T_b \approx 3.0 \times 10^{8} \, \mathrm{K}$ and the ocean-crust boundary moves to $\rho_t \approx 3.6\times 10^{9}\, \mathrm{g \ cm^{-3}}$. The outburst heating rate is $L_{\rm nuc} \approx 2.7 \times 10^{35} \, \mathrm{ergs \ s^{-1}}$ and the quiescent luminosity of the neutron star is $L_{\rm q} \approx 8.9 \times 10^{33} \, \mathrm{ergs \ s^{-1}}$. As a consequence, the ocean remains $T_b \gtrsim 1.0 \times 10^{8} \, \mathrm{K}$ for $\approx 700 \, \mathrm{days}$ into quiescence and takes $\approx 1400 \, \mathrm{days}$ to cool completely.

We also run a model of the \xte \ outburst \citep{fridriksson2010, fridriksson2011}. The model uses a neutron star mass of $M=1.6\, \mathrm{M_{\odot}}$, a neutron star radius of $R=11.6\,{\rm km}$, a core temperature of $T_{\rm core}=3 \times 10^7 \, \mathrm{K}$, an impurity parameter of $Q_{\rm imp} = 7$, and a $Q_{\rm shallow} = 0.17 \, \MeV$ per accreted nucleon shallow heat source, which agrees with the quiescent light curve fit from \citet{turlione2015}. During the accretion outburst, the ocean temperature reaches $T_b \approx 3.1 \times 10^{8} \, \mathrm{K}$ and the ocean-crust boundary moves to $\rho_t \approx 3.9\times 10^{9}\, \mathrm{g \ cm^{-3}}$. The outburst heating rate is $L_{\rm nuc} \approx 2.1 \times 10^{36} \, \mathrm{ergs \ s^{-1}}$ and the quiescent luminosity of the neutron star is $L_{\rm q} \approx 1.9 \times 10^{34} \, \mathrm{ergs \ s^{-1}}$. As a consequence, the ocean remains $T_b \gtrsim 1.0 \times 10^{8} \, \mathrm{K}$ for $\approx 1700 \, \mathrm{days}$ into quiescence and takes $\approx 3100 \, \mathrm{days}$ to cool completely.

We also run a series of models with various time-averaged accretion rates to investigate the possible \gmode\ spectrum in other Z-sources. We run models with time-averaged accretion rates during a Z-track cycle $\langle \dot{M} \rangle / \dot{M}_{\rm Edd} = 0.1, 0.25, 0.5, 0.75$, and $1.0$, for a $1 \, \mathrm{year}$ outburst to approximate a typical Z-source outburst. We use a fiducial neutron star mass $M=1.4\, \mathrm{M_{\odot}}$ and radius $R=10\,{\rm km}$, with a core temperature $T_{\rm core}=3 \times 10^7 \, \mathrm{K}$ for all models. Models with time-averaged accretion rates between $\langle \dot{M} \rangle = 0.1 \textrm{--} 0.5\, \dot{M}_{\rm Edd}$ have ocean temperatures between $T_b \approx 4.5\times 10^{7} \textrm{--} 9.9 \times 10^7 \, \mathrm{K}$ and ocean-crust transition densities between $\rho_t \approx 7.5\times10^6 \textrm{--} 8.1\times10^7 \, \mathrm{g \ cm^{-3}}$, as can be seen in Figure~\ref{figure.transients}. Models with time-averaged accretion rates between $\langle \dot{M} \rangle = 0.5\textrm{--}1.0\, \dot{M}_{\rm Edd}$ have ocean temperatures between $T_b \approx 8.1 \times 10^7 \textrm{--} 1.4 \times 10^{8} \, \mathrm{K}$ and ocean-crust transition densities between $\rho_t \approx 8.1\times10^7 \textrm{--} 2.3\times10^8 \, \mathrm{g \ cm^{-3}}$. 

\section{the evolution of ocean $\lowercase{g}$-modes} \label{section.gmode}

The ocean \gmodes \ have been suggested as a possible origin for the observed Z-source NBOs \citep{bildsten1995} due to their characteristic frequencies between $\approx 5 \textrm{--} 7 \, \mathrm{Hz}$. The thermal \gmodes \ are non-radial oscillations, where the buoyancy arises from entropy gradients in the ocean \citep{mcdermott1983,bildsten1995}. Ocean \gmodes \ may be excited as angular momentum is transported into the ocean by a spreading boundary layer \citep{inogamov1999,inogamov2010}, as is the case with acoustic modes in the neutron star envelope \citep{philippov2016}. The \gmode \ spectrum can be analytically approximated \citep{bildsten1995} by 
\begin{eqnarray} \label{eq.gmode}
f_{n,l} \approx 2.8 \, \mathrm{Hz} \, \left[\frac{l(l+1)}{6} \frac{T_b}{3 \times 10^7 \, \mathrm{K}} \frac{56}{A} \right]^{1/2} \left(\frac{10 \, \mathrm{km}}{R} \right) \nonumber \\
\times \left[1+ \left(\frac{3n\pi}{2}\right)^2 \left( \mathrm{ln} \left\{\frac{\rho_{t}}{1 \, \mathrm{g \ cm^{-3}}}\right\} \right)^{-2} \right]^{-1/2} \ ,
\end{eqnarray}

\noindent where $n$ is the number of nodes in the ocean, $l$ is the angular wavenumber, and $R$ is the radius of the neutron star. The upper boundary of the ocean is taken at the base of the envelope near a mass density of $\rho \approx 1 \, \mathrm{g \ cm^{-3}}$, however, the frequency spectrum is largely insensitive to the location of this upper boundary. Even though Equation~\ref{eq.gmode} is derived for an isothermal ocean, the oscillation spectrum is primarily set by the temperature at the base of the ocean in non-isothermal models \citep{bildsten1998}, and Equation~\ref{eq.gmode} gives frequencies accurate within $\approx 10 \, \%$ \citep{bildsten1995}. 

Here we examine the predicted fundamental $n=1$ $l=2$ \gmode \ frequency during and after an accretion outburst. Although the $l=1$ should be more easily observed \citep{bildsten1995}, the mode frequencies are smaller by a factor of $\sqrt{3}$ and are inconsistent with observed NBO frequencies. It is unclear why the fundamental $n=1$ $l=2$ mode may be excited preferentially with respect to the $l=1$ mode. Furthermore, this calculation is only valid for slowly rotating neutron stars with $\nu_{\rm spin} \ll 300\, \mathrm{Hz}$ and with magnetic fields $B \lesssim 10^{11} \, \mathrm{G}$; these limits are consistent with the spin frequencies for the atoll sources inferred from kHz QPOs (e.g., KS~1731-260; \citealt{wijnands_vanderklis_1997}); though this method for determining $\nu_{\rm spin}$ is debated (e.g., \citealt{mendez2007}).  For simplicity, we assume that the Z-sources fall within these limits. Note, however, that at spin frequencies of $\nu_{\rm spin} \gtrsim f_{n,l}$ the ocean \gmodes \ become highly modified by the rapid rotation of the neutron star and observed NBO frequencies would then only be consistent with \gmodes \ of high radial order \citep{bildsten1996}. 

\begin{figure}
\centering
\includegraphics[width=1.0\columnwidth]{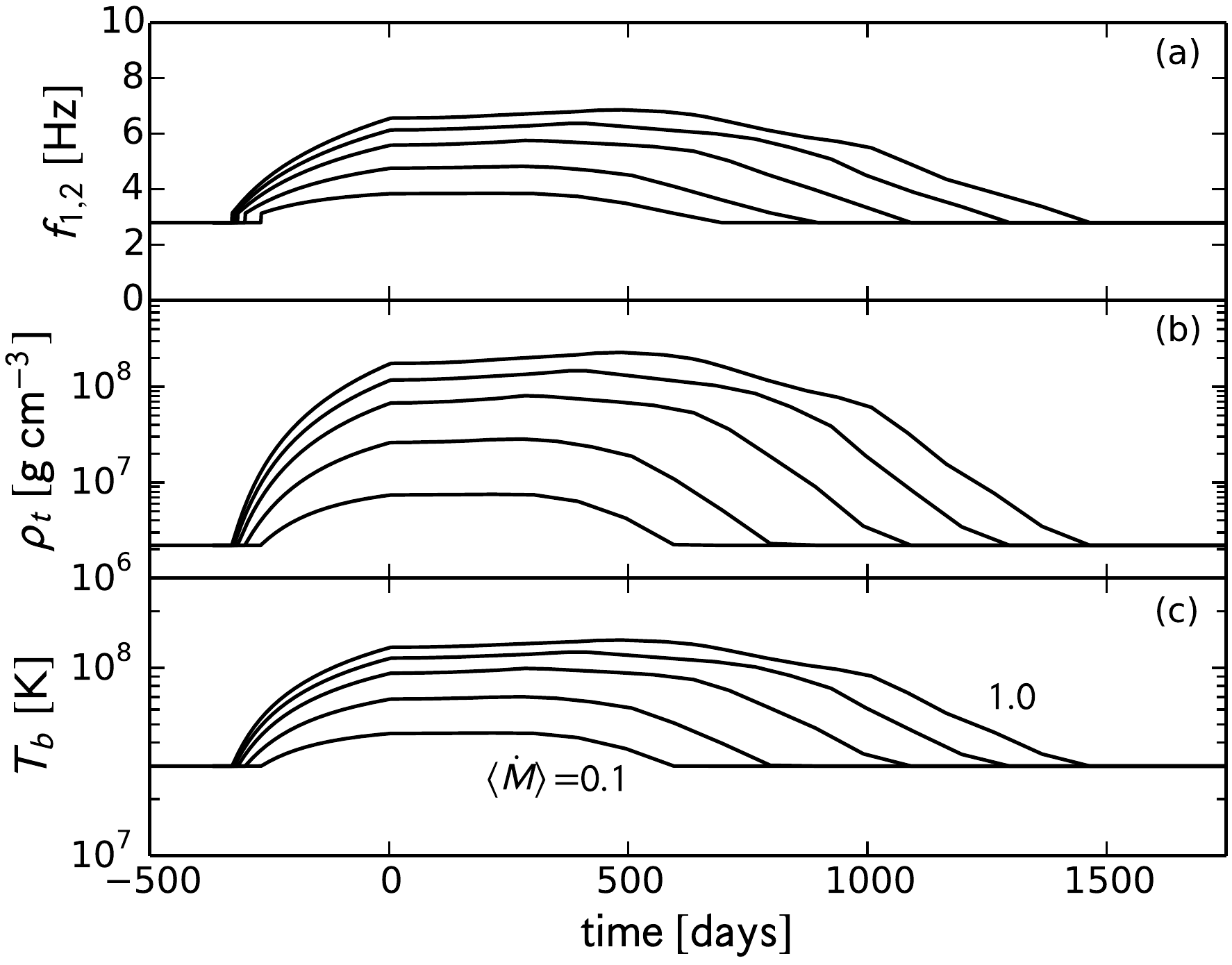}
\vspace{-0.3cm}
\caption{Thermal evolution of the ocean as a function of time during outburst/quiescence for accreting models with time-averaged accretion rates $\langle \dot{M} \rangle/ \dot{M}_{\rm Edd} = 0.1, 0.25, 0.5, 0.75$, and $1.0$ {Panel (a)}: fundamental $n=1$, $l=2$ \gmode \ frequency in the ocean (Equation~4). {Panel (b)}: the ocean-crust transition density (Equation~2). {Panel (c)}: temperature at the ocean-crust transition. \label{figure.transients}
}
\end{figure}

In a cold ocean ($T_b \approx T_{\rm core} \sim 10^7 \, \mathrm{K}$), the predicted fundamental $n=1$ $l=2$ \gmode \ frequency is $\approx 3 \, \mathrm{Hz}$. An increase in the predicted \gmode \ fundamental frequency is apparent in all Z-sources with $\langle \dot{M} \rangle \gtrsim 0.1\, \dot{M}_{\rm Edd}$ when the ocean reaches steady state. For example, sources with time-averaged accretion rates $\langle \dot{M} \rangle \lesssim 0.5 \, \dot{M}_{\rm Edd}$ have predicted fundamental \gmode \ frequencies between $\approx 3\textrm{--}5\, \mathrm{Hz}$ and sources with $\langle \dot{M} \rangle \gtrsim 0.5 \, \dot{M}_{\rm Edd}$ have predicted frequencies between $\approx 3\textrm{--}7\, \mathrm{Hz}$. Note that $\approx 7 \, \mathrm{Hz}$ is the maximum $n=1$, $l=2$ frequency reached in a model with $\langle \dot{M} \rangle = 1.0 \,\dot{M}_{\rm Edd}$, as can be seen in Figure~\ref{figure.transients}.

The predicted \gmodes \ in sources that require shallow heating, however, naturally reach frequencies $\gtrsim 7 \, \mathrm{Hz}$, as can be seen in Figure~\ref{figure.maxi}. For example, during the \maxi \ outburst, the predicted fundamental \gmode \ frequency quickly becomes $\gtrsim 7 \, \mathrm{Hz}$ after $\approx 20 \, \mathrm{days}$ and reaches $\approx 16 \, \mathrm{Hz}$ within $\approx 100$ days. The predicted fundamental remains at $\approx 16 \, \mathrm{Hz}$ for the duration of the outburst and persists at $\approx 16 \, \mathrm{Hz}$ for $\approx 500 \, \mathrm{days}$ into quiescence. The model of the \mxb \ outburst reaches frequencies $\gtrsim 7 \, \mathrm{Hz}$ within the first $\approx 30 \, \mathrm{days}$ of the outburst. The predicted fundamental \gmode \ frequency approaches $\approx 10 \, \mathrm{Hz}$ when the ocean temperature reaches a steady state temperature about $\approx 700 \, \mathrm{days}$ into the outburst.  The model of the \xte \ outburst behaves similarly to \mxb, reaching a predicted fundamental \gmode \ frequency near $\approx 10 \, \mathrm{Hz}$ when the ocean temperature reaches steady-state. The ocean cools relatively slowly in \xte, due to its larger impurity parameter of $Q_{\rm imp} = 7$, taking $\approx 3100 \, \mathrm{days}$ to cool completely \textrm{---} much longer than \maxi \ and \mxb. 

\section{discussion} \label{section.discussion}

In this study, we investigate the thermal evolution of the transient Z-sources to characterize the temporal evolution of the ocean \gmodes \ during outburst and quiescence. Before accretion begins, the predicted fundamental \gmode \ is initially $\approx 3 \, \mathrm{Hz}$ in the cold ocean ($T_b \approx T_{\rm core} \sim 10^7 \, \mathrm{K}$). During active accretion, the predicted fundamental \gmode \ frequency increases as unavoidable accretion-driven nuclear heating raises the ocean's temperature. For example, sources accreting between $\langle \dot{M} \rangle = 0.1\textrm{--}1.0\, \dot{M}_{\rm Edd}$ have predicted fundamental \gmodes \ between $\approx 3\textrm{--}7 \, \mathrm{Hz}$. Predicted fundamental \gmode \ frequencies $\gtrsim 7 \, \mathrm{Hz}$ require active shallow heating during outburst to reach ocean temperatures $\gtrsim 10^8 \, \mathrm{K}$ required to support these frequencies. Therefore, observing NBOs $\gtrsim 7 \, \mathrm{Hz}$ in transients that require shallow heating, such as \maxi, \mxb, and \xte, provides an observational test to link NBOs and the ocean \gmodes. In particular, during steady-state accretion, \maxi \ has predicted fundamental \gmode \ frequencies between $\approx 8 \textrm{--} 16\, \mathrm{Hz}$, and \mxb \ and \xte \ have predicted \gmode \ frequencies between $\approx 8 \textrm{--} 10\, \mathrm{Hz}$ \textrm{---} frequencies only possible in the hotter oceans found in sources with shallow heating. We predict that MAXI~J0556-332 had a fundamental \gmode \ frequency near $\approx 11 \, \mathrm{Hz}$ at the time of its recent activity \citep{jin2016,negoro2016,russell2016}. If shallow heating is active during the current outburst in \mxb \ \citep{negoro2015}, we predict that this source will have a fundamental \gmode \ frequency near $\approx 9 \, \mathrm{Hz}$ at the time of the most recent observation \citep{bahramian2016} after being in outburst for $\approx 160 \, \mathrm{days}$. 

NBO frequencies observed in the last outburst from \xte \ \citep{fridriksson2010} are consistent with ocean \gmodes. NBOs were observed within the first $\approx 10 \, \mathrm{weeks}$ near $\approx 7 \, \mathrm{Hz}$ \citep{homan2007}, and our model predicts fundamental \gmode \ frequencies near $\approx 7 \, \mathrm{Hz}$ after $10 \, \mathrm{weeks}$ of active accretion. Furthermore, once the ocean reaches steady state, our model of the \xte \ outburst contains predicted ocean \gmodes \ between $\approx 7 \textrm{--} 10 \, \mathrm{Hz}$, which is consistent with the observed $\approx 7\textrm{--}9 \, \mathrm{Hz}$ NBOs observed in this source during outburst \citep{homan2010}. 

Although current instrumentation may not allow observations of NBOs in \mxb \ and \maxi \ during quiescence, renewed accretion outbursts in both objects may allow observations of NBOs. We predict that observed NBOs during these sources current accretion outbursts should be $\gtrsim 5 \, \mathrm{Hz}$ in \mxb \ and $\gtrsim 10\, \mathrm{Hz}$ in \maxi. The renewed activity in \maxi \ \citep{negoro2016,russell2016} perhaps holds the best prospects for catching large NBO frequencies because the ocean in \maxi \ had not cooled to the core temperature before the most recent accretion outburst, and the predicted ocean \gmode \ frequency is already $\gtrsim 7 \, \mathrm{Hz}$. 

Note that the above observational tests for the presence of ocean \gmodes \ does not require a complete picture of how X-ray emission is modulated by ocean oscillations, which is outside the scope of this work. Uncertainties remain in the nature of the accretion flow near the neutron star surface and how the accretion flow may interact with the neutron star's outer layers. For example, a spreading boundary layer of accreted material may extend into the ocean \citep{inogamov1999,inogamov2010}, and may transport angular momentum therein. A coupling of the boundary layer and the ocean in this way may only modulate X-ray emission near the equator, however, where modes may even interfere with accretion disk emission \citep{bildsten1996}. If observations support a \gmode \ origin for the NBOs, this will motivate future work to model the ocean's coupling to a spreading boundary layer. 

If NBOs are indeed ocean \gmodes, observed NBO frequencies $\gtrsim 7 \, \mathrm{Hz}$ can then be used as a diagnostic of the shallow heating strength. For example, as shown in this work, ocean \gmode \ frequencies near $\approx 10\, \mathrm{Hz}$ ($\approx 16 \, \mathrm{Hz}$) are found in sources in steady state with $Q_{\rm shallow} \approx 1 \, \mathrm{MeV}$ ($Q_{\rm shallow} \approx 6 \, \mathrm{MeV}$). It is worth noting, however, that during outburst the ocean in \maxi \ approaches a maximum temperature set by neutrino emission around $T_b \approx 2 \times 10^9 \, \mathrm{K}$ at the depth of the shallow heat source \citep{deibel2015}. Because the ocean temperature is near maximum, a \gmode \ frequency of $\approx 16 \, \mathrm{Hz}$ is likely the largest $n=1$ $l=2$ \gmode \ that can be supported in Z-source oceans. Therefore, NBO frequencies of $\approx 16 \, \mathrm{Hz}$ only give the lower limit $Q_{\rm shallow} \gtrsim 6 \, \mathrm{MeV}$ for the shallow heating strength.

The depth of the ocean-crust transition during outburst is relevant to the study of the shallow heating mechanism. For example, the ocean-crust interface in \maxi \ moves to densities between $\rho_t \sim 10^{9} \textrm{--} 10^{11} \, \mathrm{g \ cm^{-3}}$ during outburst. This density range is characteristic of the location of the shallow heating inferred from quiescent light curves of quasi-persistent transients \citep{brown2009}, which is needed at mass densities near $\rho_{\rm shallow} \lesssim 3 \times 10^{10} \, \mathrm{g \ cm^{-3}}$. For example, shallow heating is required between $\rho_{\rm shallow} \sim 5 \times 10^9 \textrm{--} 3\times 10^{10} \, \mathrm{g \ cm^{-3}}$ in \maxi, as determined from its quiescent light curve \citep{deibel2015}. This suggests that the shallow heating mechanism is connected to the ocean-crust phase transition, and work in this direction is ongoing.

Often the observed $\approx 5\textrm{--}7\, \mathrm{Hz}$ NBOs blend into $\approx 10-20\, \mathrm{Hz}$ flaring branch oscillations (hereafter FBOs); for example, a continuous blending was observed in SCO~X-1 \citep{casella2006}. The timescale of observed NBO-FBO blending disfavors a \gmode \ origin for the FBOs. The thermal time at the depth of the ocean-crust transition (Figure~\ref{fig:thermal_time}) determines the timescale for variations in the \gmode \ frequencies. For \gmode\ frequencies $\lesssim 7 \, \mathrm{Hz}$, variations may occur on timescales of hours; variations in frequencies $\gtrsim 7\, \mathrm{Hz}$ occur on timescales of days \textrm{---} much longer than the timescale of the observed blending. The NBO-FBO blending phenomena, however, is consistent with a superposition of ocean \gmodes \ and ocean interface waves. The interface modes were determined to have frequencies of $\sim 200 \, \mathrm{Hz}$ when calculated with a rigid crust boundary condition \citep{mcdermott1988}, but may have lower frequencies near $\sim 20 \, \mathrm{Hz}$ when the flexibility of the crust is taken into account \citep{piro2005a}. It is unclear why oscillation energy might transition between the different modes, but the coupling of the \gmodes \ to the interface modes is worthy of future study. 

Quasi-periodic oscillations between $\approx 5 \textrm{--} 7 \, \mathrm{Hz}$ in the atoll sources ($L_{\mathrm{X}} \sim 0.1 \textrm{--} 0.5 \, {L_{\rm X, Edd}}$) are also consistent with ocean \gmodes. For example, the $\approx 7\, \mathrm{Hz}$ oscillation observed in 4U~1820-30 \citep{wijnands1999a,belloni2004} is consistent with a $n=1$ $l=2$ \gmode \ in a lighter ocean. In a lighter ocean, such as those considered in \citet{bildsten1995}, \gmode \ frequencies are larger by a factor of $\sim (56/16)^{1/2}$ compared to those studied in this work. This may explain the observed oscillations in 4U~1820-30, where accretion from the helium companion star \citep{wijnands1999a} would result in a different ocean composition than the one considered here.

\acknowledgements

A.D. thanks Andrew Cumming and Edward Brown for insightful comments on the manuscript. A.D. thanks Jeroen Homan for useful discussions on the nature of Z-sources and Bob Rutledge for the encouragement to publish work on the neutron star ocean. Support for A.D. was provided by the National Aeronautics and Space Administration through \emph{Chandra} Award Number TM5-16003X issued by the \emph{Chandra X-ray Observatory} Center, which is operated by the Smithsonian Astrophysical Observatory for and on behalf of the National Aeronautics and Space Administration under contract NAS8-03060. A.D. is also supported by the National Science Foundation under grant No. AST-1516969 and the Michigan State University College of Natural Science Dissertation Completion Fellowship. This material is based upon work supported by the National Science Foundation under grant No. PHY-1430152 (Joint Institute for Nuclear Astrophysics - Center for the Evolution of the Elements). 

\bibliographystyle{apj}

\begin{thebibliography}{}

\bibitem[Bahramian et al. (2016)]{bahramian2016} Bahramian, A., Heinke, C. O., Wijnands, R., \& Degenaar, N. 2016, The Astronomer's Telegram, 8699, 1

\bibitem[Bardeen \& Petterson (1975)]{bardeen1975} Bardeen, J. M., \& Petterson, J. A. 1975, ApJ, 195, L65

\bibitem[Belloni et al. (2004)]{belloni2004} Belloni, T., Parolin, I., \& Casella, P. 2004, A\&A, 423, 969

\bibitem[Bildsten \& Cumming (1998)]{bildsten1998} Bildsten, L., \& Cumming, A. 1998, ApJ, 506, 842

\bibitem[Bildsten \& Cutler (1995)]{bildsten1995} Bildsten, L., \& Cutler, C. 1995, ApJ, 449, 800

\bibitem[Bildsten et al. (1996)]{bildsten1996} Bildsten, L., Ushomirsky, G., \& Cutler, C. 1996, ApJ, 460, 827

\bibitem[{Bisnovaty{\u i}-Kogan} \& {Chechetkin} (1979)]{bisnovatyi1979} {Bisnovaty{\u i}-Kogan}, G.~S., \& {Chechetkin}, V.~M. 1979, Soviet Physics Uspekhi, 22, 89

\bibitem[Brown (2015)]{dstar} Brown, E. F. 2015, dStar: Neutron star thermal evolution code, Astrophysics Source Code Library, ascl:1505.034

\bibitem[Brown et al. (1998)]{brown1998b} Brown, E. F., Bildsten, L., \& Rutledge, R. E. 1998, ApJ, 504, L95

\bibitem[Brown \& Cumming (2009)]{brown2009} Brown, E. F., \& Cumming, A. 2009, ApJ, 698, 1020

\bibitem[Cackett et al. (2010)]{cackett2010} Cackett, E. M., Brown, E. F., Cumming, A., Degenaar, N., Miller, J. M., \&
Wijnands, R. 2010, ApJ, 722, L137

\bibitem[Cackett et al. (2008)]{cackett2008} Cackett, E. M., Wijnands, R., Miller, J. M., Brown, E. F., \& Degenaar, N. 2008, ApJ, 687, L87

\bibitem[Casella et al. (2006)]{casella2006} Casella, P., Belloni, T., \& Stella, L. 2006, A\&A, 446, 579

\bibitem[Degenaar et al. (2009)]{degenaar09} Degenaar, N., Wijnands, R., Wolff, M. T., et al. 2009, MNRAS, 396, L26

\bibitem[Degenaar et al. (2014)]{degenaar2014} Degenaar, N., Medin, Z., Cumming, A., et al. 2014, ApJ, 791, 47

\bibitem[Degenaar et al. (2015)]{degenaar2015} Degenaar, N., Wijnands, R., Bahramian, A., et al. 2015, MNRAS, 451, 2071

\bibitem[Deibel et al. (2015)]{deibel2015} Deibel, A., Cumming, A., Brown, E. F., \& Page, D. 2015, ApJ, 809, L31

\bibitem[Dubus et al. (2004)]{dubus2004} Dubus, G., Kern, B., Esin, A. A., Rutledge, R. E., \& Martin, C. 2004, MNRAS, 347, 1217

\bibitem[Farouki \& Hamaguchi (1993)]{farouki93} Farouki, R. T., \& Hamaguchi, S. 1993, Phys. Rev. E, 47, 4330

\bibitem[Fridriksson et al. (2015)]{fridriksson2015} Fridriksson, J. K., Homan, J., \& Remillard, R. A. 2015, ApJ, 809, 52

\bibitem[Fridriksson et al. (2010)]{fridriksson2010} Fridriksson, J. K., Homan, J., Wijnands, R., et al. 2010, ApJ, 714, 270

\bibitem[Fridriksson et al. (2011)]{fridriksson2011} Fridriksson, J. K., Homan, J., Wijnands, R., et al. 2011, ApJ, 736, 162

\bibitem[Gupta et al. (2007)]{gupta2007} Gupta, S., Brown, E. F., Schatz, H., M{\"o}ller, P., \& Kratz, K.-L. 2007, ApJ,
662, 1188

\bibitem[Haensel \& Zdunik (1990)]{haensel1990} Haensel, P., \& Zdunik, J. L. 1990, A\&A, 227, 431

\bibitem[Haensel \& Zdunik (2003)]{haensel2003} Haensel, P., \& Zdunik, J. L. 2003, A\&A, 404, L33

\bibitem[Haensel \& Zdunik (2008)]{haensel2008} Haensel, P., \& Zdunik, J. L. 2008, A\&A, 480, 459

\bibitem[Hasinger \& van der Klis (1989)]{hasinger1989} Hasinger, G., \& van der Klis, M. 1989, A\&A, 225, 79

\bibitem[Hertz et al. (1992)]{hertz1992} Hertz, P., Vaughan, B., Wood, K. S., et al. 1992, ApJ, 396, 201

\bibitem[Homan (2012)]{homan2012} Homan, J. 2012, ApJ, 760, L30

\bibitem[Homan et al. (2015)]{homan2015} Homan, J., Fridriksson, J. K., \& Remillard, R. A. 2015, ApJ, 812, 80

\bibitem[Homan et al. (2014)]{homan2014} Homan, J., Fridriksson, J. K., Wijnands, R., Cackett, E. M., Degenaar, N., Linares, M., Lin, D., \& Remillard, R. A. 2014, ApJ, 795, 131 

\bibitem[Homan et al. (2011)]{homan2011} Homan, J., Linares, M., van den Berg, M., \& Fridriksson, J. 2011, The Astronomer's Telegram, 3650, 1

\bibitem[Homan et al. (2010)]{homan2010} Homan, J., van der Klis, M., Fridriksson, J. K., et al. 2010, ApJ, 719, 201

\bibitem[Homan et al. (2007)]{homan2007} Homan, J. van der Klis, M., Wijnands, R., et al. 2007, ApJ, 656, 420

%\bibitem[Horowitz et al. (2010)]{horowitz2010} Horowitz, C. J., Schneider, A. S., \& Berry, D. K. 2010, Phys. Rev. Lett., 104, 231101

\bibitem[Inogamov \& Sunyaev (1999)]{inogamov1999} Inogamov, N. A., \& Sunyaev, R. A. 1999, Astronomy Letters, 25, 269

\bibitem[Inogamov \& Sunyaev (2010)]{inogamov2010} Inogamov, N. A., \& Sunyaev, R. A. 2010, Astronomy Letters, 36, 848

\bibitem[Jin \& Kong (2016)]{jin2016} Jin, R., \& Kong, A. K. H. 2016, The Astronomer's Telegram, 8530, 1

\bibitem[Jonker et al. (2002)]{jonker2002} Jonker, P. G., van der Klis, M., Homan, J., et al., 2002, MNRAS, 333, 665

\bibitem[Kuulkers et al. (1994)]{kuulkers1994} Kuulkers, E., van der Klis, M., Oosterbroek, T., et al., 1994, A\&A, 289, 795

\bibitem[Lense \& Thirring (1918)]{lense1918} Lense, J., \& Thirring, H. 1918, Physikalische Zeitschrift, 19, 156

%\bibitem[Lewin \& van der Klis (2006)]{lewin2006} Lewin, W. H. G., \& van der Klis, M. 2006, Compact Stellar X-ray Sources

\bibitem[Lin et al. (2009)]{lin2009} Lin, D., Remillard, R. A., \& Homan, J. 2009, ApJ, 696, 1257

\bibitem[Matsumura et al. (2011)]{matsumura2011} Matsumura, T., Negoro, H., Suwa, F., et al. 2011, The Astronomer's Telegram, 3102, 1

\bibitem[McDermott et al. (1988)]{mcdermott1988} McDermott, P. N., van Horn, H. M., \& Hansen, C. J. 1988, ApJ, 325, 725

\bibitem[McDermott et al. (1983)]{mcdermott1983} McDermott, P. N., van Horn, H. M., \& Scholl, J. F. 1983, ApJ, 268, 837

\bibitem[M\'{e}ndez \& Belloni (2007)]{mendez2007} M\'{e}ndez, M., \& Belloni, T. 2007, MNRAS, 381, 790

\bibitem[Negoro et al. (2015)]{negoro2015} Negoro, H., Furuya, K., Ueno, S., et al., 2015, The Astronomer's Telegram, 7943, 1

\bibitem[Negoro et al. (2016)]{negoro2016}  Negoro, H., Nakajima, M., Fujiwara, T., et al., 2016, The Astronomer's Telegram, 8513, 1

\bibitem[Page \& Reddy (2013)]{page2013} Page, D., \& Reddy, S. 2013, Physical Review Letters, 111, 241102 

\bibitem[Paxton et al. (2011)]{paxton2011} Paxton, B., Bildsten, L., Dotter, A., Herwig, F., Lesaffre, P., \& Timmes, F.
2011, ApJS, 192, 3

\bibitem[Paxton et al. (2013)]{paxton2013}  Paxton, B., Cantiello, M., Arras, P., et al. 2013, ApJS, 208, 4

\bibitem[Paxton et al. (2015)]{paxton2015} Paxton, B., Marchant, P., Schwab, J., et al.  2015, ApJS, 220, 15

\bibitem[Philippov et al. (2016)]{philippov2016} Philippov, A. A., Rafikov, R. R., \& Stone, J. M. 2016, ApJ, 817, 62

\bibitem[Piro \& Bildsten (2005)]{piro2005a} Piro, A. L., \& Bildsten, L. 2005, ApJ, 619, 1054

\bibitem[Potekhin \& Chabrier (2000)]{potekhin2000} Potekhin, A. Y., \& Chabrier, G. 2000, Phys. Rev. E, 62, 8554

\bibitem[Russell \& Lewis (2016)]{russell2016} Russell, D. M., \& Lewis, F. 2016, The Astronomer's Telegram, 8517

%\bibitem[Rutledge et al. (2002)]{rutledge2002} Rutledge, R. E., Bildsten, L., Brown, E. F., Pavlov, G. G., Zavlin, V. E., \&
%Ushomirsky, G. 2002, ApJ, 580, 413

\bibitem[Sato (1979)]{sato1979} Sato, K. 1979, Progress of Theoretical Physics, 62, 957

\bibitem[Steiner (2012)]{steiner2012} Steiner, A. W. 2012, Phys. Rev. C, 85, 055804

\bibitem[Stella \& Vietri (1998)]{stella1998} Stella, L., \& Vietri, M. 1998, ApJ, 492, L59

\bibitem[Strohmayer (1993)]{strohmayer1993} Strohmayer, T. E. 1993, ApJ, 417, 273

\bibitem[Sugizaki et al. (2013)]{sugizaki2013} Sugizaki, M., Yamaoka, K., Matsuoka, M., et al. 2013, PASJ, 65, 58

\bibitem[Titarchuk et al. (2014)]{titarchuk2014} Titarchuk, L., Seifina, E., \& Shrader, C. 2014, ApJ, 789, 98

\bibitem[Turlione et al. (2015)]{turlione2015} Turlione, A., Aguilera, D. N., \& Pons, J. A. 2015, A\&A, 577, A5 

\bibitem[van der Klis et al. (1996)]{vanderklis1996} van der Klis, M., Swank, J. H., Zhang, W., et al., 1996, ApJ, 469, L1

\bibitem[Wijnands et al. (2002)]{wijnands2002} Wijnands, R., Guainazzi, M., van der Klis, M., \& Me\'{e}ndez, M. 2002, ApJ,
573, L45

\bibitem[Wijnands et al. (2004)]{wijnands2004} Wijnands, R., Homan, J., Miller, J. M., \& Lewin, W. H. G. 2004, ApJ, 606,
L61

\bibitem[Wijnands et al. (2001)]{wijnands2001} Wijnands, R., Miller, J. M., Markwardt, C., Lewin, W. H. G., \& van der
Klis, M. 2001, ApJ, 560, L159

\bibitem[Wijnands et al. (2003)]{wijnands2003} Wijnands, R., Nowak, M., Miller, J. M., Homan, J., Wachter, S., \& Lewin,
W. H. G. 2003, ApJ, 594, 952

\bibitem[Wijnands et al. (1999)]{wijnands1999a} Wijnands, R., van der Klis, M., \& Rijkhorst, E.-J. 1999, ApJ, 512, L39

\bibitem[Wijnands \& van der Klis (1997)]{wijnands_vanderklis_1997} Wijnands, R. A. D., \& van der Klis, M. 1997, ApJ, 482, L65

\bibitem[Wijnands et al. (1997)]{wijnands1997} Wijnands, R. A. D., van der Klis, M., Kuulkers, E., Asai, K., \& Hasinger, G. 1997, A\&A, 323, 399

\end{thebibliography}

\end{document}